\documentclass[1p,authoryear,11pt,twocolumn]{elsarticle}

    \makeatletter
    \def\ps@pprintTitle{%
      \let\@oddhead\@empty
      \let\@evenhead\@empty
      \def\@oddfoot{\reset@font\hfil\thepage\hfil}
      \let\@evenfoot\@oddfoot
    }
    \makeatother

\usepackage[centertags]{amsmath}
\usepackage{pifont}
\usepackage{geometry}
\usepackage{fleqn}
\usepackage{natbib}
\usepackage{color}
\usepackage{amssymb}
\usepackage{epsfig}
\usepackage{rotating}
\usepackage{verbatim}
\usepackage{amsfonts}
\usepackage[margin=10pt,font=small,labelfont=bf]{caption}

\usepackage{pifont}
\usepackage{subfigure}

\usepackage{graphicx}
\usepackage{subfigure}
\usepackage[centertags]{amsmath}

\usepackage{wrapfig}
\newcommand{\be}{\begin{equation}}
\newcommand{\ee}{\end{equation}}

\DeclareCaptionLabelFormat{adja-page}{\hrulefill\\#1 #2 \emph{(previous page)}}

\begin{document}
\title{Co-evolution of resource trade-offs driving species interactions in a host-parasite network: An exploratory model}
\author[rvt]{C.F.˜ M$^\text{c}$Quaid\corref{cor1}}
%\ead{cfm21@bath.ac.uk}
\author[rvt]{N.F.˜Britton}
\cortext[cor1]{Corresponding author\\Tel.: +44 1225 38 5669\\Email: cfm21@bath.ac.uk }
\address[rvt]{Department of Mathematical Sciences, University of Bath, Claverton Down, Bath, BA2 7AY, UK}
\begin{abstract}
Patterns of nestedness and specialization asymmetry, where specialist species interact mainly with generalists while generalists interact with both generalists and specialists, are often observed in mutualistic and antagonistic bi-partite ecological networks. These have been explained in terms of the relative abundance of species, using a null model that assigns links in proportion to abundance, but doubts have been raised as to whether this offers a complete explanation. In particular, host-parasite networks offer a variety of examples in which the reverse patterns are observed.

We propose that the link between specificity and species-richness may also be driven by the co-evolution of hosts and parasites, as hosts allocate resources to optimize defence against parasites, and parasites to optimize attack on hosts. In this hypothesis, species interactions are a result of resource allocations. This novel concept, linking together many different arguments for network structures, is introduced through the adaptive dynamics of a simple ecological toy system of two hosts and two parasites.

We analyse the toy model and its functionality, demonstrating that co-evolution leads to specialization asymmetry in networks with closely related parasites or fast host mutation rates, but not in networks with more distantly related species. Having constructed the toy model and tested its applicability, our model can now be expanded to the full problem of a larger system.
\end{abstract}
\maketitle

\section{Introduction}
\label{Introduction}
What drives the association between the host-specificity of parasites and the parasite species-richness of the hosts that they infect? Generalist parasites are often found in hosts with both high and low parasite species diversity, while specialist parasites are found mainly in hosts with a rich diversity of parasites \citep{P97,VPKS05}. This association, between the specificity of parasites and the species-richness of the hosts that they infect, is known as specialization asymmetry \citep{VPKS05}, or in a slightly stronger case as nestedness, when those species occurring in a species-poor assemblage form a non-random subset of those assemblages with a higher species richness  \citep{PG00}. These patterns can be seen from parasites in fish species  \citep{P97} to fleas and their hosts  \citep{VPKS05}, although there is some debate on the extent to which this is evident, and many counterexamples exist  \citep{P97,P07,VPPJ01}.

Nestedness is also evident in many other food webs \citep{IMBBBDEFJJLLLOVWW09}, and there is mixed evidence for whether the addition of parasites to these webs should increase or decrease their relative nestedness \citep{HS08,LDK06}. Nestedness is a feature which is particularly prevalent in mutualistic networks \citep{BJMO03,VA03,VA04}, an interesting fact given the complete reversal in interaction types involved when compared to host-parasite networks. Nevertheless, in order to fit parasites into food web models, it is important to understand the forces behind such structural traits as nestedness, particularly as antagonistic networks are instead generally expected to be compartmentalized (\citealt{B10,TF08,T05}, but see \citealt{FMVFW11} for the effects of scale-dependence).

What is clear is that structural patterns of species interactions in ecological networks, such as nestedness and anti-nestedness, are not random \citep{JMSSP10}. \citet{VPKS05} constructed a null model based on host abundance to account for such patterns, yet there may be many more explanations for the link between the specificity of a parasite and the parasite species-richness of the hosts that it infects \citep{LPJBO06,P07}. This does not, for example, explain the frequent occurrence of anti-nestedness. Further explanations for nestedness in networks include complementarity \citep{RJB07}, based on phenotypic matching between species, and competitive load \citep{BFPFLB09}, based on a new species entering a network targeting a host with less competition provided by resident parasite species.

Another possible driving force behind the link between specificity and species richness is related to the levels of defence that a host exhibits. For example, avian fleas with different levels of specificity target hosts with different levels of T-cell mediated immune response \citep{MCG05}. Generalist parasites target hosts with weak levels of immune response, while parasites with fewer host species exploit those with both strong and weak immune responses. This also has an effect on the parasite species-richness of host species, with hosts with stronger immune responses being parasitized by a greater number of species. In this instance, then, the specificity of parasites and specialization asymmetry appears to be related to the host immune response \citep{MCG05}. Here, this idea is turned around slightly, but the concept of a relationship between host response and specificity is maintained.

It is becoming increasingly apparent that the co-evolution of species is an important driving force in host-parasite relationships \citep{BWB09}. It is also widely acknowledged that trade-offs in resource allocation are responsible for much evolutionary drive in parasites, such as the link between virulence and transmission\citep{MA83}. In our paper it is postulated that the link between specificity and richness may be driven by the co-evolution of both hosts and parasites when balancing the allocation of resources (see \citealt{PM04}). These resources are devoted in different degrees to interactions with one species versus another, either for infection or for preventing infection.

As an example, the influenza virus binds to cell-surface oligosaccharides via a sialic acid receptor. The receptor type may have one of two conformations: Neu 5Ac\(\alpha\)(2,3)-Gal or Neu5Ac\(\alpha\)(2,6)-Gal. A host species may have one of the two or both \citep{CPD10}, but the virus must adapt to one linkage type at the expense of the other. Other examples include phenotypic trait matching, such as the shape of mouthparts of ectoparasites \citep{GHBGU09} or nectar holder size and shape in mutualistic pollinator networks \citep{VBCC09}. From the host perspective, a behavioural example involves reindeer and other herd animals, which often group in a reaction to parasitism by biting flies, reducing their chances of attack. This does, however, lead to an increased risk of exposure to other pathogens which rely on host density for transmission \citep{H88}.

Using the above concepts, the following is proposed: if an infection is more prevalent, then a host will have a higher likelihood of coming into contact with it and adapting to fight it, allocating a greater amount of resource to this and increasing its immune response. A host has, however, limited resources with which to do so \citep{PBH03,PM04}. Similarly, a parasite may infect multiple hosts. It will, however, be better adapted to infect some than others, and again there will be a trade-off in terms of its efficiency in infecting a host species (see \citealt{P98}). It is therefore assumed that both host and parasite species trade off their resources between those species that they target. This trade-off aims to incorporate all of the ideas discussed above. A more abundant species will provide more available hosts for a parasite. A lower competitive load will encourage infection of that host as an untapped resource. Lastly, complementarity will ensure that a species will target another with complimentary trait values (parasites will infect hosts which are more vulnerable to them, while hosts will protect themselves against parasites which are more of a threat).

\begin{figure}
\centering
\subfigure[]{
\label{specass1} 
\includegraphics[width=1.5cm]{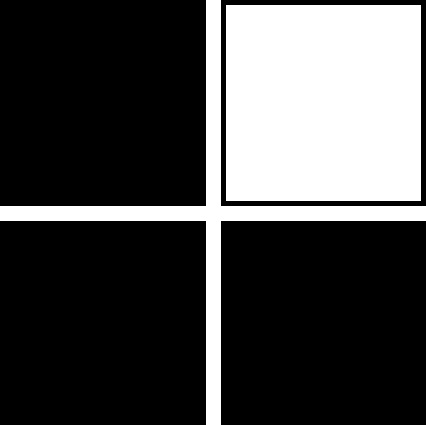}}
\subfigure[]{
\label{specass2} 
\includegraphics[width=1.5cm]{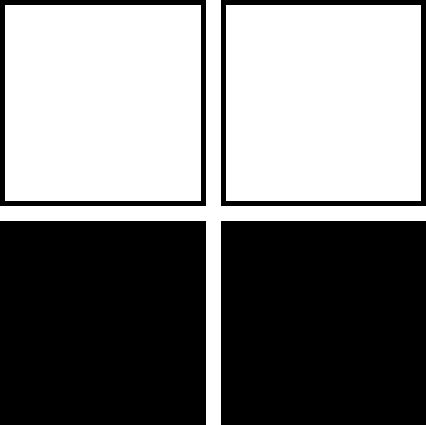}}
\subfigure[]{
\label{specass3} 
\includegraphics[width=1.5cm]{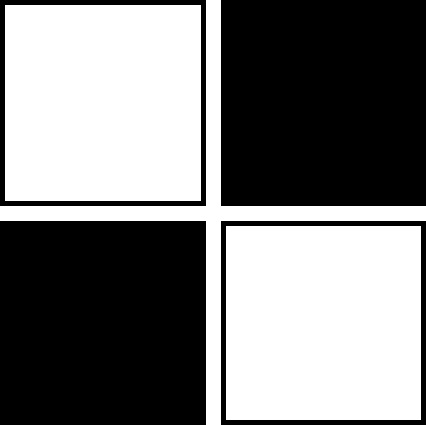}}
\subfigure[]{
\label{specass4} 
\includegraphics[width=1.5cm]{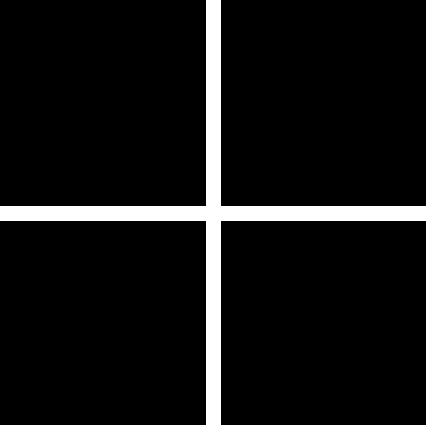}}
\caption{Possible interaction matrices for a 2-host (y-axes) 2-parasite (x-axes) network. A black square represents an interaction between two species and a white no interaction. Plots demonstrate specialization asymmetry: (a) a generalist and a specialist parasite and (b) two specialist parasites in the same host, and specialization symmetry: (c) two specialist parasites in opposite hosts and (d) two generalist parasites. }
\label{specass}
\end{figure}

With this in mind, we create a toy model here which investigates the co-evolution of trade-offs in a dynamical manner, for a four-species system containing two species of each type; hosts and parasites. This may be thought of as a cluster of species forming a `compartment' in a larger food web, and hence more general results may be inferred from the results obtained (see \citealt{JBSSP09}). Although this is simply a toy problem, figure \ref{specass} gives an indication of what specialization asymmetry might look like in this case, and hence the patterns we might expect the model to show for nestedness and anti-nestedness in a larger system. It is important to stress that in order to fully understand the influence of these trade-offs on nestedness, the investigation of a larger system is necessary (in prep.). 

We outline a model describing the system and its mathematical motivation, following this with a brief analysis and an investigation of the co-evolutionary dynamics. The results are compared to a previous null model based on abundance \citep{VPKS05}, in order to demonstrate the compatibility of this theory with our model.

\section{Model}
A standard susceptible-infected system is assumed, with two species of hosts and two of parasites. This model has the potential for both specialist and generalist parasites (in one or both hosts respectively) as well as species-poor and species-rich hosts (containing neither parasite, one only or both). \(S_i\) refers to susceptibles of host type \(i\), while \(I_{ij}\) refers to infecteds of host type \(i\) with parasite type \(j\), where in this instance \(i,j\in\mathbb{N}_2\). We then have
\begin{align}
\frac{dS_i}{dt} = &\alpha_iN_i - \sum_{j\in\mathbb{N}_2} a_{ij}c_{ij} S_i F^i_j-\omega_i S_iN_i,\notag\\
\frac{dI_{ij}}{dt}= &a_{ij}c_{ij} S_iF^i_j- \gamma_{ij}I_{ij}-\omega_iI_{ij} N_i.
\label{Model}
\end{align}

The model contains birth (\(\alpha_i\)) and death (\(\omega_i\)) rates dependent on the host species \(i\), as well as infection-related death; death rate \(\gamma_{ij}\) of host species \(i\) due to parasite species \(j\). \(N_i=S_i+I_{i1}+I_{i2}\) represents the total population size of host species \(i\). 

The {\em maximum} force of infection $F_j^i$ of parasite species $j$ on host species $i$ is given by \[F_j^i=\sum_{k\in\mathbb{N}_2}\beta_j^{ik}I_{kj},\] where $\beta_j^{ik}$ is the pairwise potential infectious contact rate for the transfer of parasite $j$ from host $k$ to host $i$.  In our model, the {\em actual} force of infection $G_j^i$ is assumed to be moderated by the strategies adopted by the parasite $j$ and the host $i$.  It is given by
\begin{equation}
G_j^i = a_{ij}c_{ij}F_j^i
  = a_{ij}c_{ij}(\beta_j^{i1}I_{1j} + \beta_j^{i2}I_{2j}),\notag
\end{equation}
with $0 \leq a_{ij} \leq 1$, $0 \leq c_{ij} \leq 1$. Here $a_{ij}$ is a parasite-related trait defining the relative probability of success of parasite $j$'s attack on host $i$, and $c_{ij}$ is a host-related trait defining the relative probability of failure of host $i$'s defence against parasite $j$.  All else being equal, parasites benefit from values of $a_{ij}$ that are as high as possible, while hosts benefit from values of $c_{ij}$ that are as low as possible.  

We assume, however, that each parasite species $j$ has a fixed amount of resource to allocate to infection, and that there is therefore a trade-off between the strength $a_{1j}$ of its attack against host $1$ and the strength $a_{2j}$ of its attack against host 2. This trade-off is assumed to be a decreasing function, which is species specific and is not dependent upon the population or environment. A host species $i$, meanwhile, varies strategy \(c_{ij}\) in order to reduce transmission of parasite $j$, and a similar trade-off is presumed.

Transmission of infection to a susceptible host then depends on the actual force of infection, a measure of both the propensity of the parasite to infect that host, as well as the host's propensity to defend itself against the parasite.

In reality, it is difficult to determine the shapes such trade-offs take \citep{BWB09}. A general trade-off shape therefore allows for a greater understanding of different possible evolutionary outcomes \citep{K06}. Points at which an evolutionary stable strategy (ESS) may exhibit evolutionary branching also often depend on the nature of the trade-off function presumed; more specifically, whether it is concave or convex, and the extent of this \citep{K06}. Levins' fitness set approach \citep{ RVM06} has previously influenced intuitive thoughts on the effects of these shapes under normal evolutionary conditions, with a generalist expected if a trade-off is convex, or weak, and either specialist expected if a trade-off is concave, or strong. 

The trade-off shape is determined here by a species-specific power ($\theta_j$ for parasite species $j$ and $\phi_i$ for host species $i$), where
\[a_{2j}=(1-a_{1j}^{\theta_j})^{\frac{1}{\theta_j}} \quad \text{ and } \quad c_{i2}=(1-c_{i1}^{\phi_i})^{\frac{1}{\phi_i}}.\] 

\begin{figure}
\centering
\includegraphics[width=8.4cm]{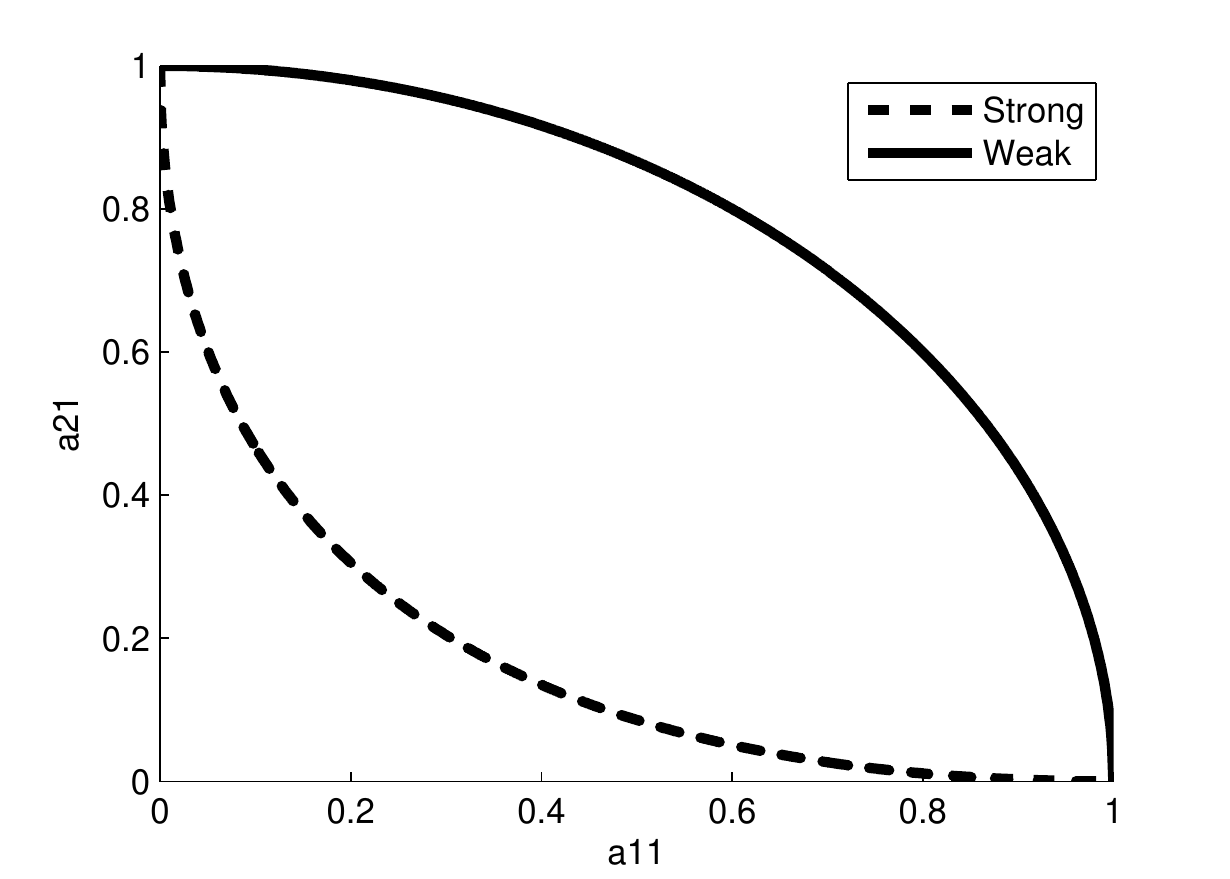}
\caption{Examples of trade-off shapes for trait values of parasite species 1, where the trade-off is either strong ($\theta_1=0.5$) or weak ($\theta_1=2$)}
\label{Tradeoff}
\end{figure}

For parasite species \(j\), \(\theta_j<1\) implies a strong trade-off, and \(\theta_j>1\) implies a weak trade-off (see figure \ref{Tradeoff}). A parasite is a perfect generalist if \(a_{1j}=a_{2j}=(0.5)^{\frac{1}{\theta_j}}\), which is henceforth termed the {\em neutral point}. It is a complete specialist if \(a_{1j}=0\) and \(a_{2j}=1\), or \textit{vice versa}. This is similar for host species. Note that the tendency is for trade-offs to be strong, not weak \citep{RVM06}.

The second derivative of the trade-off function, the degree to which it is concave or convex, can have an enormous effect on the possibility of evolutionary branching (see \citealt{K06}). The sign of this depends wholly upon the value of the species-specific trade-off shape, and so the system will exhibit very different behaviour around this point. The more concave a function is, the more likely the system is to exhibit branching (Kisdi 2006). Note that the second derivative is maximized (either positive or negative depending on the trade-off shape) for parasite $j$ at \(a_{1j}=(0.5)^{\frac{1}{\theta_j}},\) where \[\frac{d^2 a_{2j}}{da^2_{1j}} = (\theta_j-1),\] and similarly for host species. This affects the direction in which trait-values mutate, which will be discussed in greater detail in the following sections.
\section{Analysis}
This model is analysed using adaptive dynamics. This assumes rare mutants with marginally different phenotypic effects to residents, which may then invade the population if their growth rate is positive in an equilibrium environment \citep{DIR03}. This approach does assume clonal reproduction, but the results still hold for random mating in monomorphic diploid populations with polygenic traits if mutations are rare with small phenotypic effect \citep{LBF01,RVM06}. The growth rate of a mutant phenotype while rare is termed the invasion fitness, and is important in determining whether that mutant may invade, and potentially replace, a resident \citep{DD04}. The population evolves in the direction of the fitness gradient (the change in the invasion fitness with respect to change in the mutant trait value) as successive mutations occur and then spread through the population \citep{DD04}. The invasion fitness can then be used to discover singularities, where the fitness gradient of any local mutant is zero, and to investigate the nature of these singularities \citep{GKMM98}. An example of the derivation of the invasion condition is given below for parasite 1 in a system without co-infection or recovery; the method is similar for the second parasite species. 
% In this case, if a mutant is initially able to invade the population then it is presumed to replace the resident, an assumption which does not hold near to a bifurcation point in the population dynamics, or an evolutionary equilibrium point (Law et al. 2001)
\subsection{Parasite invasion conditions}
\label{growthpara}
In determining the invasion fitness of a mutant parasite in the population, here a mutant \(I_{i1}'\) with strategy (\(a_{11}',a_{21}'\)) of parasite 1, with the resident population at a stable, non-trivial equilibrium (\(S_i=S_i^*,\) \(I_{ij}=I_{ij}^*\)), the linearized dynamics of the mutant are expressed as follows:
\begin{align}
&\frac{dI'_{11}}{dt}=a_{11}'c_{11}S^*_1F^{1'}_1-\gamma_{11}I'_{11}-\omega_1 I_{11}' N^*_1,\notag\\
&\frac{dI'_{21}}{dt}=a_{21}'c_{21}S^*_2F^{2'}_1-\gamma_{21}I'_{21}-\omega_2 I_{21}' N^*_2.
\end{align}

In the full system of equations for the dynamics, including the resident dynamics, the resident is at a stable steady equilibrium. The Jacobian of the system can be therefore split into the original dynamics and a separate submatrix determined by the mutant, given by \(\boldsymbol A_1\).

 Defining \(\delta_{11}=a_{11}'-a_{11}\) and \(\delta_{21}=a_{21}'-a_{21}\), both small, we get
\begin{equation}
 \boldsymbol A_{1} =\left( \begin{array}{cc}
\boldsymbol A_1(1,1) & \boldsymbol A_1(1,2)\\
\boldsymbol A_1(2,1) & \boldsymbol A_1(2,2)
\end{array} \right),
\label{P1jac}
\end{equation}
%\[ A_{1} =\left( \begin{array}{cc}
%\delta_{11}c_{11}S^*_1\beta^{11}_1-a_{11}c_{11}S^*_1\beta^{12}_1\frac{I_{21}^*}{I_{11}^*} & (\delta_{11}+a_{11})c_{11}S^*_1\beta^{12}_1\\
%(\delta_{21}+a_{21})c_{21}S^*_2\beta^{21}_1 & \delta_{21}c_{21}S^*_2\beta^{22}_1-a_{21}c_{21}S^*_2\beta^{21}_1\frac{I_{11}^*}{I_{21}^*}
%\end{array} \right).\] 
where the equilibrium of the system is used to guarantee that \[I_{i1}^*(\gamma_{i1}+\omega_i N_i^*)=a_{i1}c_{i1}S_i^*F_1^{i*},\]
and hence
\begin{align}
&\boldsymbol A_1(n,n)I^*_{n1}=\delta_{n1}c_{n1}S_n^*\beta^{nn}_1I^*_{n1}-a_{n1}c_{n1}S^*_{n}\beta^{nn-1}_1I_{n-11}^*,\notag\\
&\boldsymbol A_1(n,n-1)=(\delta_{n1}+a_{n1})c_{n1}S_n^*\beta^{nn-1}_1.
\end{align}

The stability of \(\boldsymbol A_1\) is then used to indicate the potential for invasion. As it is difficult to interpret anything useful from the eigenvalues, a sign-equivalent proxy can be found for the growth rate of the mutant by investigating the trace and determinant of \(\boldsymbol A_1\). The determinant is given by 
\[I_{11}^*I_{21}^*\det(\boldsymbol A_{1})=-c_{11}c_{21}S^*_1S^*_2 u_{1}(\delta_{11},a_{11},\delta_{21},a_{21}),\]
where, to leading order of \(\delta\),
\begin{align}
u_1(\delta_{11},a_{11},\delta_{21},a_{21})=&\beta^{12}_1 \beta^{21}_1 I^*_{11}I^*_{21} (\delta_{11}a_{21}+a_{11}\delta_{21})\notag\\
&+\beta^{21}_1\beta^{11}_1 (I^*_{11})^2 \delta_{11}a_{21}\notag\\
&+ \beta^{12}_1 \beta^{22}_1(I^*_{21})^2 a_{11}\delta_{21}.
\label{P1cond}
\end{align}

When \(a_{11}'=a_{11}\), then tr(\(\boldsymbol A_1)<0\) and \(\det(\boldsymbol A_1)=0\), as \(\delta_{n1}=0\). For the small perturbations resulting from mutation we can therefore rely on the determinant condition for stability analysis, as the trace will remain negative. The cases for $I_{11}^*=0$ and $I_{21}^*=0$ are discussed later. The determinant will be negative, and hence the system is unstable and the mutant invades, if 
\begin{equation} 
u_1(\delta_{11},a_{11},\delta_{21},a_{21})>0
\end{equation}

Interpreting this condition, invasion of a mutant parasite can be seen to be driven by the following; the first term is due to inter-species transmission, and is minimized at the neutral point \(a_{11}=a_{21}\), hence driving towards generalism. This is due to an increased invasion probability when a mutant is investing more equally than the resident, as \((\delta_{11}a_{21}+a_{11}\delta_{21})\) is larger. When \(a_{11}=a_{21}\) then this term is zero, and no possible mutant can invade due to this. When \(a_{11}<(0.5)^{\frac{1}{\theta_1}}\), this implies that \(a_{21}>(0.5)^{\frac{1}{\theta_1}}\). Hence we require a mutant with a larger value of \(a_{11}\) to promote invasion, and \textit{vice versa} for \(a_{11}>(0.5)^{\frac{1}{\theta_1}}\). So this term stabilizes at and attracts to the neutral point, promoting generalism.

The final two terms compare the use of the two available hosts. For example, if transmission due to host 1 (\(\beta^{21}_1\beta^{11}_1 (I_{11}^*)^2\)) is high, then a mutant with larger \(a_{11}'\) will invade, ensuring host 1 is utilized. These terms demonstrate an increased invasion potential if the mutant increases infection of the species with higher infection rates (\(\beta^{21}_1\beta^{11}_1\) compared to \(\beta^{12}_1\beta^{22}_1\)) and on which the resident relies more (whether \(I^*_{11}\) is greater or less than \(I^*_{21}\)), as \(\delta_{11}\) and \(\delta_{21}\) have opposite signs. This ensures that the mutant is spread as much as possible to susceptible hosts.

Note that the case where either \(I_{11}^*=0\) or \(I_{21}^*=0\) sees the parasite shy away from a completely protected host. These points are only obtainable, for a four species system where the maximum force of infection is not zero, if the actual force of infection is zero, i.e. $a_{11}c_{11}=0$ or $a_{21}c_{21}=0$ as appropriate. For the case where \(I_{11}^*=0\), the submatrix of the Jacobian is given by 
\begin{equation}
\boldsymbol A_1=\left( \begin{array}{cc}
-\gamma_{11}-\omega_{11}(S_1^*+I_{12}^*) & 0\\
(\delta_{21}+a_{21})c_{21}S^*_2\beta^{21}_1 & \delta_{21}c_{21}S^*_2\beta^{22}_1
\end{array} \right).\notag
\label{P1condc10}
\end{equation}
The mutant can thus invade only if \(\delta_{21}>0\), i.e. if \(a_{11}'<a_{11}.\) A similar situation arises for \(I_{21}^*=0\). If both \(I_{11}^*\) and \(I_{21}^*=0\) then the system is at a trivial equilibrium, in which case no mutant can invade.
\subsection{Host invasion conditions}
\label{growthhost}
Similarly to the parasite case above, a mutant population (\(S_1'\), \(I_{1j}'\)) with trait value \(c_{1j}'\) of host 1 is introduced at low densities to the resident population at equilibrium (\(S_i = S_i^*,\) \(I_{ij}=I_{ij}^*\)). The dynamics of the subsystem are given by
\begin{align}
\frac{dS'_1}{dt}=&\alpha_{1}N_1'-a_{11}c_{11}'S_1'F_1^{1*}-a_{12}c_{12}'S_1'F_2^{1*}-\omega_1S_1'N_1^*,\notag\\
\frac{dI'_{11}}{dt}=&a_{11}c_{11}'S_1'F_1^{1*}-\gamma_{11}I_{11}'-\omega_1 I_{11}' N_1^*,\notag\\
\frac{dI'_{12}}{dt}=&a_{12}c_{12}'S_1'F_2^{1*}-\gamma_{12}I_{12}'-\omega_1 I_{12}' N_1^*.
\end{align}

The equilibrium conditions are taken from equation \ref{Model}, and we define \(\epsilon_{11}=c_{11}'-c_{11}\) and \(\epsilon_{12}=c_{12}'-c_{12}\), both small. Again we investigate the  submatrix of the Jacobian, given by
\begin{equation}
 \boldsymbol C_{1} =\left( \begin{array}{ccc}
 \boldsymbol C_{1}(1,1)& \alpha_1 & \alpha_1\\
 a_{11}(\epsilon_{11}+c_{11})F_1^{1*} & -\frac{a_{11} c_{11} S_1^* F_1^{1*}}{I_{11}^*} & 0\\
 a_{12}(\epsilon_{12}+c_{12})F_2^{1*} & 0 & -\frac{a_{12}c_{12}S_1^*F_2^{1*}}{I_{12}^*}
\end{array} \right),
\label{H1jac}
\end{equation}
where \[\boldsymbol C_{1}(1,1)=\alpha_1\bigg(1-\frac{N_1^*}{S_1^*}\bigg)-a_{11}\epsilon_{11}F_1^{1*}-a_{12}\epsilon_{12}F_2^{1*}.\]

Now, as the subsystem is three-dimensional, the eigenvalues are given as solutions to an equation of the form \(\lambda^3 +b_1 \lambda^2 + b_2 \lambda + b_3\),
where
\begin{align}
b_1=&-\text{tr}(\boldsymbol C_1),\notag\\
2 b_2=&\text{tr}^2(\boldsymbol C_1)-\text{tr}(\boldsymbol C_1^2),\notag\\
b_3=&-\det(\boldsymbol C_1).\notag
\end{align}

For stability, it is required that \(b_1, b_2, b_3>0\) and \(b_1b_2>b_3\). When \(c'_{11}=c_{11}\), both \(b_1\) and \(b_2\) are positive and \(O(1)\), as \(\epsilon=0\).
%and
%\begin{align}
%2 b_2=&\boldsymbol C_1(1,1)\boldsymbol C_1(2,2)+\boldsymbol C_1(1,1)\boldsymbol C_1(3,3)\notag\\
%&+\boldsymbol C_1(2,2)\boldsymbol C_1(3,3)-\boldsymbol C_1(1,3)\boldsymbol C_1(3,1)\notag\\
%&-\boldsymbol C_1(1,2)\boldsymbol C_1(2,1),\notag
%\end{align}
%which
On the other hand,
\[b_3=-\frac{1}{I^*_{11}I^*_{12}}a_{11}a_{12}S_1^*F^{1*}_1F^{1*}_2 w_1(\epsilon_{11},c_{11},\epsilon_{12},c_{12}),\]  
 where
\begin{align}
w_1(\epsilon_{11},c_{11},\epsilon_{12},c_{12})= & c_{12}\epsilon_{11}\alpha_1I^*_{11}+c_{11}\epsilon_{12}\alpha_1 I^*_{12}\notag\\
&-c_{11} c_{12}a_{11}S^*_1 F^{1*}_1\epsilon_{11}\notag\\
&-c_{11} c_{12} a_{12} S^*_1 F^{1*}_2\epsilon_{12}.\label{GenHost1cond}
\end{align}
Now this is \(O(\epsilon)\), and is zero when \(c'_{11}=c_{11}\), hence \(b_1b_2>b_3\). The criteria for {\em instability} (and invasion) when mutations are small can therefore be reduced to \begin{equation}
w_1(\epsilon_{11},c_{11},\epsilon_{12},c_{12})>0.
\label{Hcond}
\end{equation}

Interpreting this condition, it can be seen that mutant invasion depends on a balance of terms. The first two terms from equation \ref{GenHost1cond} push the host towards a generalized defence. For example, a particularly large value of \(c_{11}\) makes the second term likely to be larger, and hence a mutant with a smaller trait value \(c'_{11}\) will invade, to ensure that \(\epsilon_{12}>0\).

The final two terms decrease in importance (\(c_{11}c_{12}\) decreases), compared to the initial terms, as the host specializes. If the pressure from parasite 1 is higher (the actual force of infection is higher) then the first of these terms will be larger. A host mutant with a smaller trait value will therefore invade, in order to make $\epsilon_{11}$ negative and the sum of the final two terms positive. In this way the mutant host protects itself against parasite 1 to a greater extent. These terms then account for the pressure that a parasite places on the host, and the host's reaction to this.

The case where either \(I_{11}^*\) or \(I_{12}^*\) is zero, which again occurs if $a_{11}c_{11}=0$ or $a_{12}c_{12}=0$ as appropriate, shows that a host will not defend itself against a non-threatening parasite, and if the system is at a trivial equilibrium, then no mutant may invade. For example, if \(I_{11}^*=0\), then equation \ref{H1jac} is given by
\begin{equation}
 \boldsymbol C_{1} =\left( \begin{array}{ccc}
\boldsymbol C_1(1,1)& \alpha_1 & \alpha_1\\
 0 & -\gamma_{11}-\omega_1N_1^*& 0\\
 a_{12}c_{12}'F_2^{1*}& 0 & -\frac{a_{12}c_{12}S_1^*F_2^{1*}}{I_{12}^*}
\end{array} \right),\notag
\label{a1=0}
\end{equation}
where $\boldsymbol C_1(1,1)= \alpha_1\bigg(1-\frac{N_1^*}{S_1^*}\bigg)-a_{12}\epsilon_{12}F_2^{1*}.$ In this case, a stronger than necessary condition is that a mutant will always invade if \(a_{12}\epsilon_{12}F^{1*}_2<0,\) i.e. if \(c_{11}'>c_{11}.\) A similar situation arises for \(I_{12}^*=0.\) 

The conditions for equilibrium can be independently verified using the next-generation tools outlined in \citet{HCD10}. For \(\rho(\boldsymbol M)\) defined to be the spectral radius of matrix \(\boldsymbol M\), and \(s(\boldsymbol M)\) the spectral bound, the Jacobian \(\boldsymbol C_1\) is decomposed such that
\begin{equation}
\boldsymbol C_1=\boldsymbol F_1-\boldsymbol V_1,\notag
\end{equation}
for \(\boldsymbol F_1\) and \(\boldsymbol V_1\) satisfying \(s(-\boldsymbol V_1)<0\), \(\boldsymbol V_1^{-1}\geq 0\) and \(\boldsymbol F_1 \geq 0\). \(\boldsymbol F_1\) and \(\boldsymbol V_1\) are taken as matrices representing the appearance and disappearance of individuals from the system respectively. The invasion condition is then given by
\begin{equation}
\rho(\boldsymbol F_1\boldsymbol V_1^{-1})>1,\notag
\end{equation}
For further details see \citet{HCD10}. In this case, we take
\begin{equation}
\boldsymbol  F_{1} =\left( \begin{array}{ccc}
 \alpha_1& \alpha_1 & \alpha_1\\
0 & 0 & 0\\
0& 0 & 0
\end{array} \right)\notag
\label{NGMHF}
\end{equation}
and
\begin{equation}
 \boldsymbol V_{1} =\left( \begin{array}{ccc}
\boldsymbol V_1(1,1)& 0 &0\\
-a_{11}(\epsilon_{11}+c_{11})F^{1*}_1 & \frac{a_{11}c_{11}S^*_{11}F_1^{1*}}{I_{11}^*} & 0\\
-a_{12}(\epsilon_{12}+c_{12})F^{1*}_2& 0 &  \frac{a_{12}c_{12}S^*_{11}F_2^{1*}}{I_{12}^*}
\end{array} \right),\notag
\label{NGMHV}
\end{equation}
where \(\boldsymbol V_1(1,1)=\alpha_1\frac{N_1^*}{S_1^*}+a_{11}\epsilon_{11}F_1^{1*}+a_{12}\epsilon_{12}F_2^{1*}\). After some straightforward algebra, this yields an identical condition to equations \ref{GenHost1cond} and \ref{Hcond}. This approach is not useful for the parasite invasion conditions, as individuals enter the system through more than one class (see \citealt{HCD10}.
%\begin{align}
%\rho(F_1V_1)&=\\
%&\frac{\alpha_1 \big(S_1^*c_{11}c_{12}+I_{11}^*c_{12}(\epsilon_{11}+c_{11})+I_{12}^*c_{11}(\epsilon_{12}+c_{12})\big)}{(\alpha_1 %N_1^*+a_{11}\epsilon_{11}F^{1*}_1S_1^*+a_{12}\epsilon_{12}F_2^{1*}S_1^*)c_{11}c_{12}}.\notag\\
%\end{align}

From the initial analysis, therefore, it appears that parasites infect more vulnerable hosts, but also aim to be generalists. Hosts trade-off between the pressures exerted by different parasite species, aiming to lower this as much as possible, but similarly aim to generalize their defence. Note that the cases for mutants of parasite and host species 2 are similar to the above. This is all as expected, and demonstrates the biological validity of our trade-off. 

Simplifications of the above scenario demonstrate that the co-evolution of all four species is vitally important, as expected. If one or more species is missing from the system then the trade-off ensures that the dynamics are trivial, with those that have two antagonists splitting their resources as before, while those that have only one concentrate their resources on that one. This demonstrates the importance of such a system, incorporating both multiple hosts and parasites, unlike many previous approaches taken when modeling host-parasite systems. 

\section{Results}
Simulation of the results follows the method of \citet{DL96}, concerning the frequency and impact of selection, where evolutionary dynamics occur at a much slower rate than population dynamics (see \citealt{DM05}. This relies on the derivative with respect to the mutant trait-value of the growth rate of the mutant in a population of residents, given by the dominant eigenvalue for each species from matrices \ref{P1jac} and \ref{H1jac}. As a result of the appropriate eigenvalue equations, discussed in sections \ref{growthpara} and \ref{growthhost}, and noting that $\lambda$ must be small, the dominant eigenvalues may be approximated up to some positive multiplicative coefficients by $-\det(A_j)$ and $\det(C_i)$ respectively. We therefore use the following general equation for species $k$ with trait-value \(s_k\) and the appropriate approximation for the eigenvalue \(E_k\):
\begin{equation}
\frac{d}{dt}s_k=f_k(s)\cdot\left.\frac{\partial}{\partial s_k'} E_k (s_k',s) \right|_{s_k'=s_k},
\label{DL}
\end{equation}
where \(f_k\) is the evolutionary rate coefficient (see \citealt{DL96}. This is calculated for an environment determined by resident trait-values $s$ for all species.

Changes to trait values with time are then investigated, where species can co-evolve. It is presumed that \(\phi_i=1\) for hosts, so the trade-off shape is linear, and initial trait values are at the neutral point (0.5 here), in order to attempt to separate the effects of hosts and parasites. There are a number of possible cases for different \(\theta\) values; trade-off curves of both parasite species may be either concave or convex (\(\theta<1\) or \(\theta>1\) respectively).

\begin{figure}
\centering
\subfigure[]{
\label{PIPs} 
\includegraphics[width=4cm]{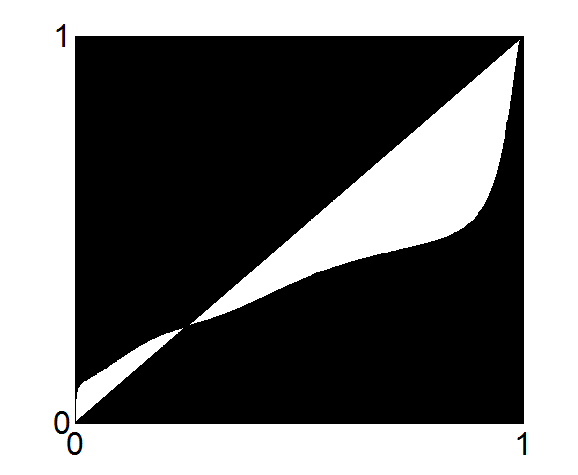}}
\subfigure[]{
\label{PIPw} 
\includegraphics[width=4cm]{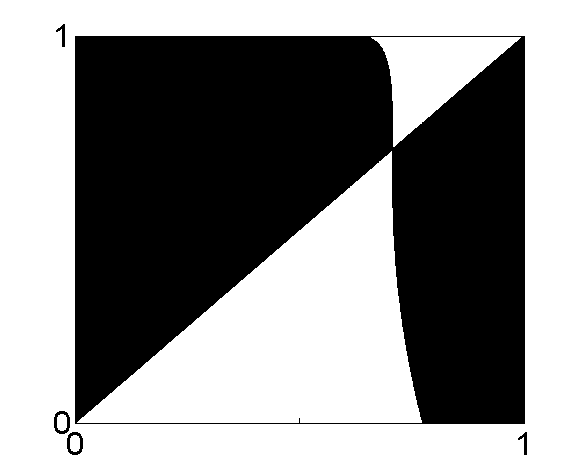}}
\caption{Pairwise invasion plots for parasite 1 in a symmetric environment, with intra-species transmission higher than inter-species transmission. Mutant trait values are on the y-axes, while resident trait values are on the x-axes. Trade-off shapes are (a) strong for the parasites and (b) weak for the parasites, while linear for the hosts.}
\label{PIP}
\end{figure}

A look at the pairwise invasion plots for cases with different trade-off strengths demonstrates the outcomes that we expect (see figure \ref{PIP}). Pairwise invasion plots (PIPs) indicate when a mutant can invade (shaded) or not (white) depending on its trait value relative to the resident trait value. This property depends on the mutant's invasion fitness at low frequency in a resident population. The population evolves as small mutations occur which move the population off the diagonal \citep{GKMM98}. If the mutation is successful (i.e. the mutant is in a shaded area) then the mutant population grows and it displaces the resident, to become the new resident \citep{GKMM98}. In figure \ref{PIPs} here, for example, this occurs until the population's trait value reaches an extreme, depending on which side of the central point the resident trait value begins. Species with a strong trade-off are likely to evolve to be specialists (figure \ref{PIPs}), while those with a weak trade-off are expected to evolve towards generalism (figure \ref{PIPw}).

\begin{figure}
\subfigure[Weak trade-off for parasites (\(\theta_j=2\))]{
\label{TvsTa} 
\includegraphics[width=8.4cm]{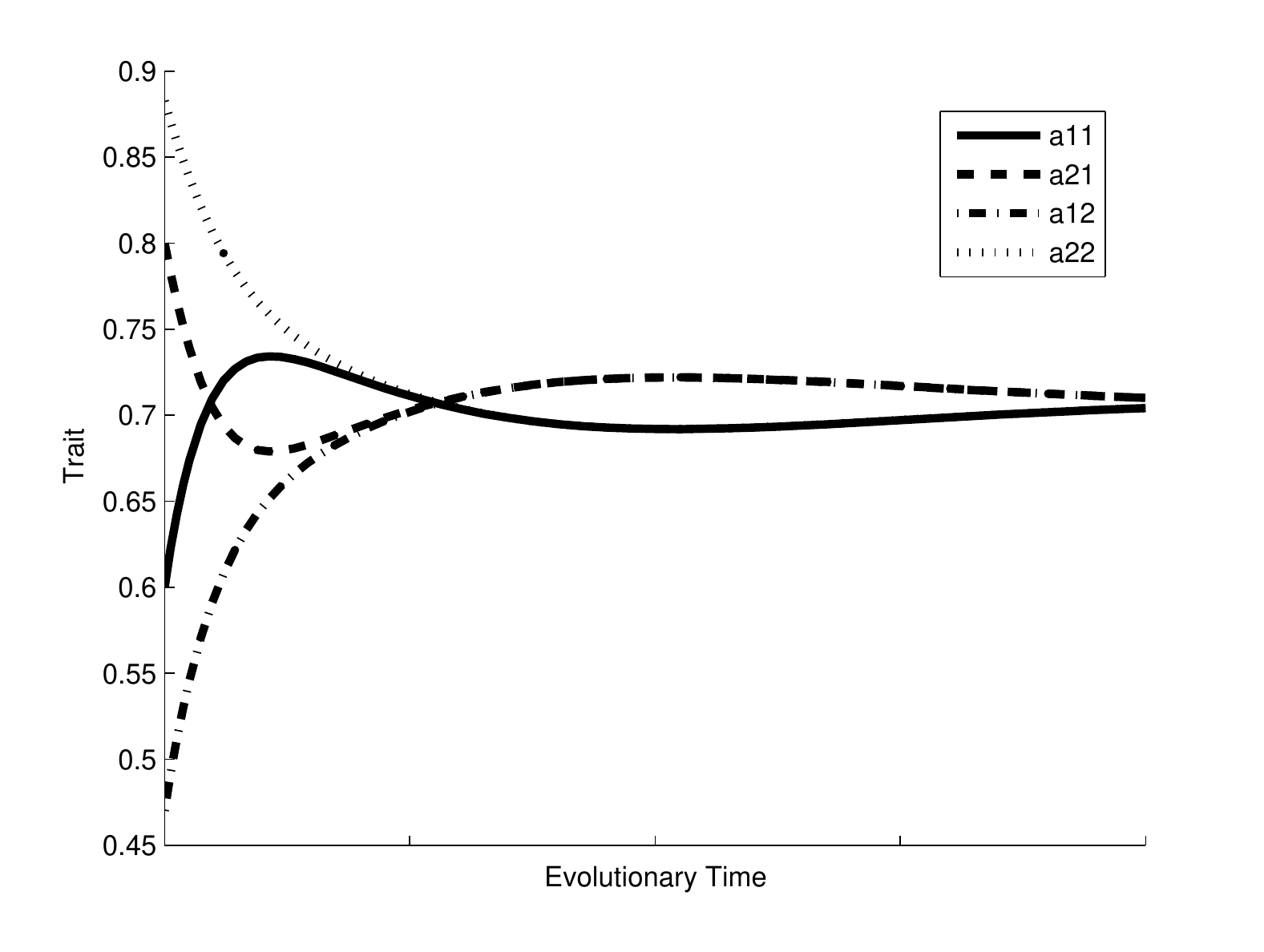}}
\subfigure[Mixed trade-offs for parasites (\(\theta_1=2,\theta_2=0.5\))]{
\label{TvsTb} 
\includegraphics[width=8.4cm]{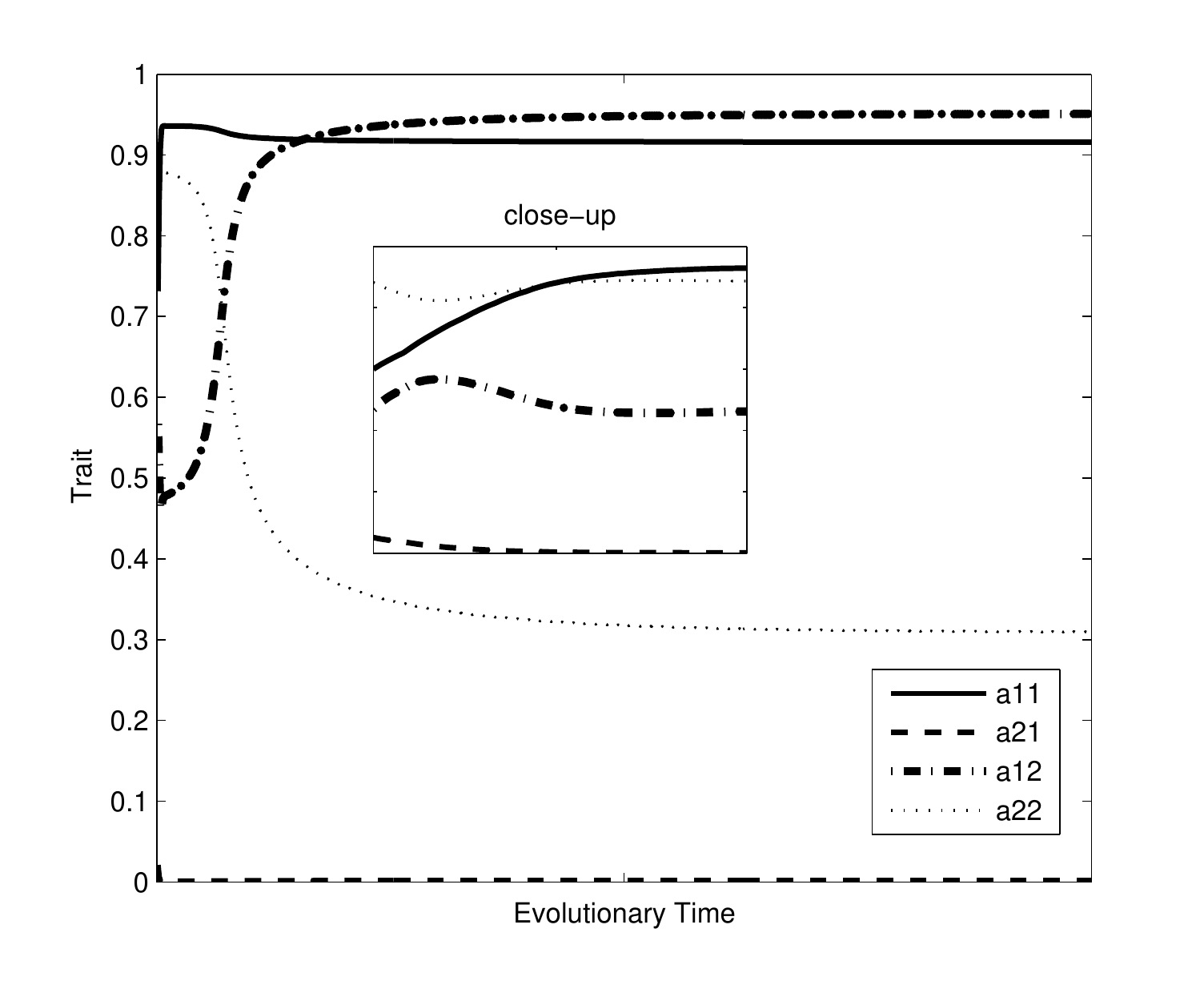}}
\caption{Examples of the evolution of trait values with time in a symmetric environment, with intra-species transmission higher than inter-species transmission. Rates of change of trait values are calculated directly from the growth rates of mutant traits in a resident population. Trade-off shapes are linear for hosts and (a) weak or (b) mixed for parasites. The inset in (b) shows the initial dynamics of the system before the slower host mutations have had an effect. Note that $a_{11}$ and $a_{21}$ represent trait values for parasite species 1 in hosts 1 and 2 respectively, while $a_{12}$ and $a_{22}$ represent the same for parasite species 2.}
\label{TvsT}
\end{figure}

\begin{figure}
\centering
\subfigure[Parasites in the same host]{
\label{TvsT2a} 
\includegraphics[width=8.4cm]{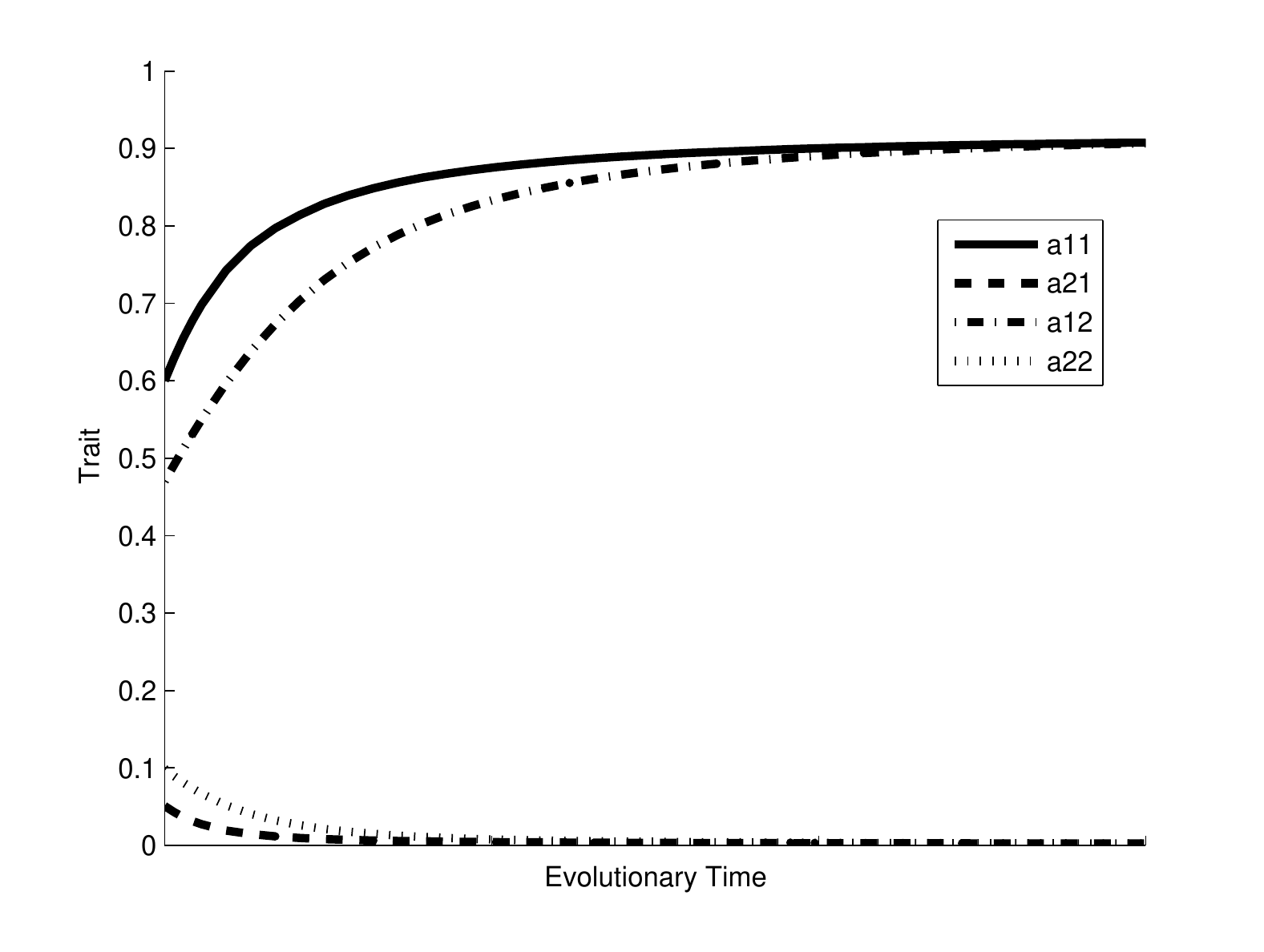}}
\subfigure[Parasites in opposite hosts]{
\label{TvsT2b} 
\includegraphics[width=8.4cm]{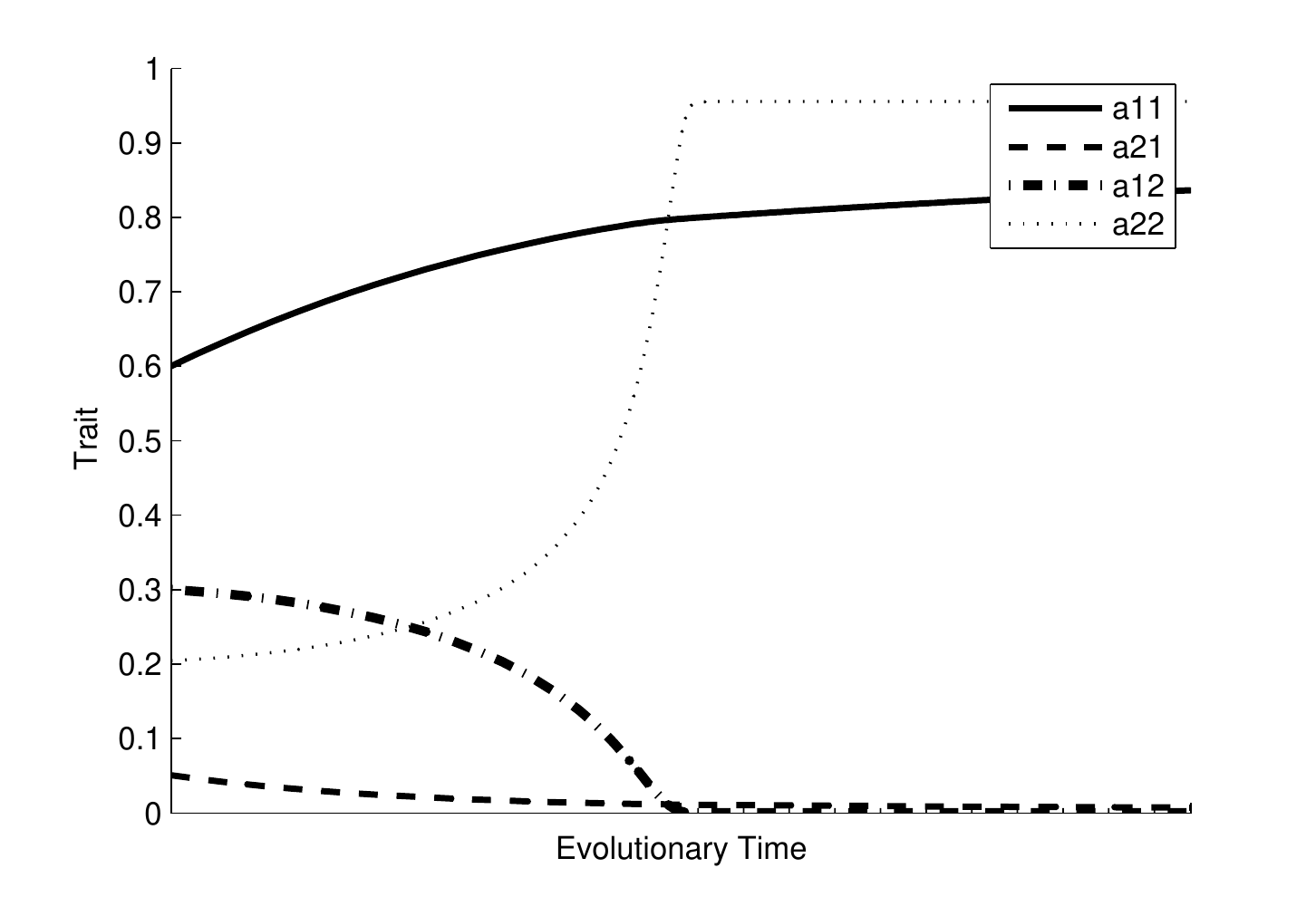}}
\caption{Examples of the evolution of trait values with time in a symmetric environment, where the trade-off shapes for both parasites are strong (\(\theta_j=0.5\)). Different endpoints occur as a result of the initial equilibrium values for susceptible and infected hosts. These result in parasites in (a) the same or (b) opposite hosts.}
\label{TvsT2}
\end{figure}

Figure \ref{PIP} demonstrates the evolution of one parasite only, in a static environment where no other species evolves. In a full analysis this will not be the case. This motivates us to follow the co-evolution of trait values for all four species simultaneously. Results are demonstrated in figures \ref{TvsT} and \ref{TvsT2}.

In figures \ref{TvsT} and \ref{TvsT2}, although mutation rates are taken to be identical for hosts and parasites, the growth rate of mutant hosts in a resident environment is significantly slower than that of parasites. The host trait values are not presented here, although their importance is discussed later.

Again we assume a linear trade-off for hosts, which have initial trait values at their respective neutral points. In figure \ref{TvsTa}, where \(\theta_1,\theta_2>1\), generalist parasites evolve. Here the trait values of parasites can be seen to evolve towards their neutral points. Note that the system cycles around the neutral point until both parasites are perfect generalists. This corresponds to figure \ref{specass4}. 

For the case where the trade-offs for the parasites are strong for one parasite and weak for the other (figure \ref{TvsTb}), we obtain the coexistence of a relative generalist and an extreme specialist in the same hosts. Note that this system takes much longer over evolutionary time to equilibrate than other cases, due to the slow rate growth rate of mutant host populations. Hence we see different dynamics over short (figure \ref{TvsTb} inset) and long (figure \ref{TvsTb}) time-scales. Before the hosts are able to react to the presence of the parasites, we see a generalist host that is more focused on the opposite host to the specialist (trait values $a_{11}$ and $a_{22}$ are higher), whereas once the host mutant populations have had an effect, we see both the relative generalist and the specialist parasite are more focused on the same host (trait values $a_{11}$ and $a_{12}$ are higher). As a host which contains only a generalist would be able to focus its defensive efforts on that parasite, and hence, as can be seen from equation \ref{H1jac}, the parasite would not target it, we see only comparative generalists here. This system demonstrates specialization asymmetry, and indeed nestedness (as far as that is plausible in such a small system) over longer evolutionary time-scales, corresponding to figure \ref{specass1}.

For the case where both \(\theta_j<1\) (figure \ref{TvsT2}), specialists always evolve. These can evolve to be in the same hosts (i.e. \(a_{11}\) and \(a_{12}\) evolve to the same extreme, figure \ref{TvsT2a}), or in the opposite host (\(a_{11}\) and \(a_{12}\) at the opposite extremes, figure \ref{TvsT2b}). These scenarios bear resemblances to specialization asymmetry (figure \ref{specass2}) and compartmentalization (figure \ref{specass3} where the network is split into separate sub-networks that are not linked to one another) respectively.

\subsection{Initial trait values}
From the parasite conditions for invasion, it can be seen that the behaviour of the cases depends heavily on the initial equilibrium conditions, which are a result of the initial trait values. The behaviour of each scenario pivots around which side of case-specific points the initial trait values lie, similarly to which side of an evolutionary stable strategy an initial point lies in a pairwise invasion plot (see figure \ref{PIP}). Each case has only a limited number of evolutionary end points for trait values, and evolves to these. An analysis of the initial trait values shows interesting results.

\begin{figure}
\centering
\subfigure[Low host mutation rate]{
\label{SS12a} 
\includegraphics[width=7cm]{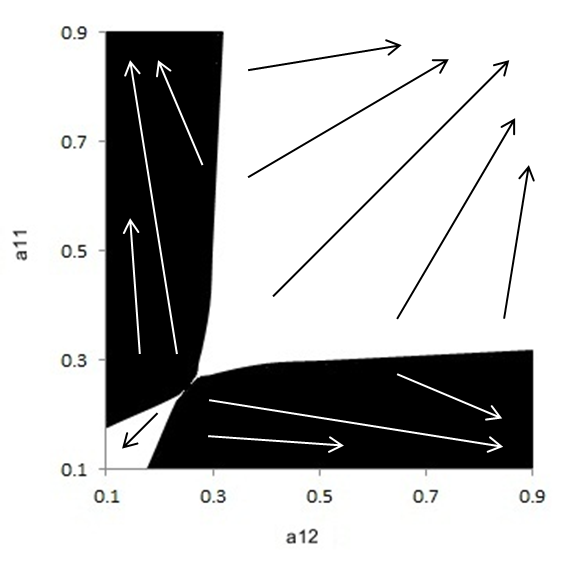}}
\subfigure[High host mutation rate]{
\label{SS12b} 
\includegraphics[width=7cm]{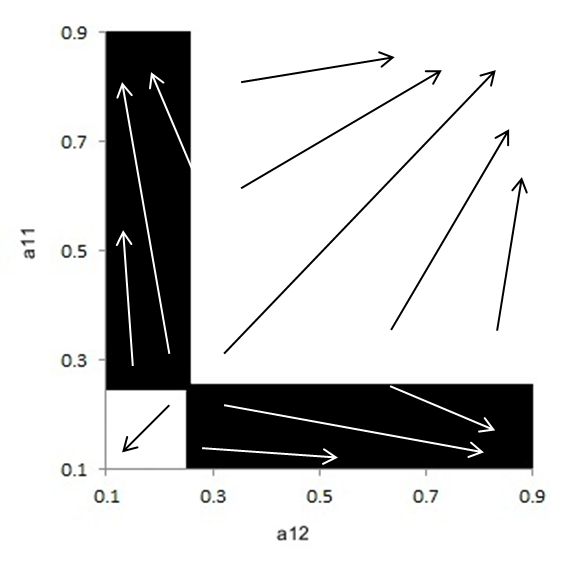}}
\caption{A sketch of the end-points of evolution, in terms of species-richness, for different initial trait values for a strong trade-off ($\theta_j=0.5$) for parasite species. Black areas denote those initial trait values that evolve to have parasites in opposite hosts, while white areas denote those for which parasites evolve to be in the same host for (a) a low host mutation rate and (b) a high mutation rate for hosts. Arrows indicate the direction of evolution of trait values. Note that the trait value for each parasite with respect to the second host is determined by the trait value for the first, and hence is not included in these plots.}
\label{SS12}
\end{figure}

For the case of a weak trade-off for both parasites (figure \ref{TvsTa}), although generalists occur almost exclusively in ``opposite'' hosts, the trait values of the parasites evolve to be so similar that only a very careful inspection can detect the difference. In other words, two generalists occur, each slightly more dependent on a different host. If one parasite has a strong trade-off and the second a weak trade-off, a relative specialist and a relative generalist occur respectively, in the same host (figure \ref{TvsTb}).This demonstrate specialization asymmetry, and occurs for all initial trait values. For the case where both parasites demonstrate a strong trade-off (figure \ref{TvsT2}), figure \ref{SS12a} shows the initial trait values which lead to specialization asymmetry.

Increasing the hosts mutation rates ($f_k$ from equation \ref{DL}) serves to alter the shape of the curves towards that found in figure \ref{SS12b}. The much faster generation time of parasites has been used in the past to justify the study of their evolution alone, as opposed to a full co-evolutionary system \citep{VA04}. Even including co-evolution, however, it has been shown by \citet{BWB09} that different mutation rates do have a significant effect. In our simulations, where the growth rates of mutant host populations in resident environments appear significantly slower than those of mutant parasite populations, increasing the mutation rates of hosts in comparison to parasites greatly increases the chances of specialization asymmetry occurring for a wider range of initial trait-values when a strong trade-off exists for parasites. This indicates not only the circumstances under which we might expect to see asymmetry, but also a possible line of experimentation to take in order to validate our results.

The addition of both recovery terms and co-infected classes to the system, computed numerically, demonstrated increased likelihoods of specialization asymmetry occurring, again dependent on both the initial trait-values and relative mutation rates of species (unpublished results).

\subsection{Abundance}
\label{Abundance}
In our toy model we included both abundance and phenotypic matching as motivators for nestedness. In order to justify the claim that the model is compatible with the concept of abundance as a driving force (see \citealt{GHBGU09,PM04,VPKS05,VBCC09}), we investigate here the manner in which hosts of different abundance influence the model outcomes. This is especially important over evolutionary, as opposed to ecological, timescales. 

There are two aspects of abundance to be compared to the model presented here. Firstly, does increased abundance of a species indeed lead to a higher number of links occurring, and secondly, does asymmetry in the assignment of links lead to nestedness? The first of these is compared to the model, while the second is a question that remains independent of the model. 

Firstly we note that, if nestedness is a result of abundance then we would expect it to occur in mutualistic and predator-prey webs, as the number of links of a species is associated solely with its abundance and not with the nature of those links. This is indeed evident (see \citealt{MPS06}), particularly in mutualistic networks, which display more nestedness than would be expected from a random, bipartite network \citep{BJMO03}. However, if this is the case then why would certain web-types display more nestedness than others (see \citealt{HS08,LDK06})? This promotes the thinking that abundance alone cannot explain patterns of nestedness in ecological networks. A further question concerns whether or not abundant species are more likely to have links simply as they are more likely to interact with other species due to their abundance, or whether this is due to the inherent benefits of interacting with a more abundant species. Our model attempts to address this, proposing that there is an evolutionary advantage to interacting with a more abundant species.

\begin{figure}
\centering
\includegraphics[width=8.4cm]{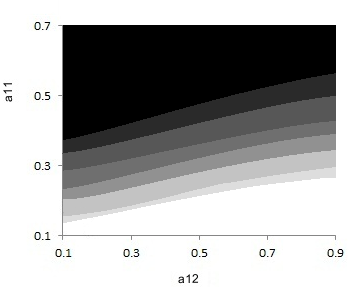}
\caption{Regions of initial trait values for both parasites which, under co-evolution with a strong trade-off, lead to parasite 1 occuring in host 1. Filled areas denote regions of initial trait values which see parasite 1 evolving to be in host 1, and succesively lighter shades denoting the increase in size of the region as the relative abundance of host species 1 increases.}
\label{abundance}
\end{figure}

For our trade-off model, note in equations \ref{P1cond} that the invasion potential of a parasite depends on the relative abundances of the different host species. The terms \(I_{11}\) and \(I_{21}\) are the combination of infection prevalence and total population size, and hence reflect that a larger population will increase invasion potential. This was also computed numerically, where it was observed that, in a symmetric environment, increasing a species' population size could drive a parasite to preferentially interact with that species (see figure \ref{abundance}). Here we look at parasite 1 interacting with host 1 and see that, although this is still dependent on initial values for traits, increasing the abundance of the host species clearly increases the likelihood of the parasite evolving to interact with that species.

\section{Discussion}
This model provides evidence for both specialization symmetry and asymmetry, but it is difficult to comment on nestedness from this. A similar model might be sufficient for larger networks, where a species-poor host could still contain more than one parasite, allowing parasites to split a host's defence. Additionally, if hosts were to trade defensive properties in an alternative manner (for example through reduced birth-rate, or the inability to reduce transmission completely to zero) this could allow generalists to exist alone in a species poor host.

There is a great deal of mixed evidence, from a number of different ecosystems under varying conditions, for nestedness in host-parasite interactions \citep{GHBGU09}. In those systems in which nestedness occurs, generalists will be in comparatively species-poor hosts, as they may occur in both species-poor and species rich hosts. In this sense the model can be related to nestedness, although its size makes a full comparison unreliable. The model also supports the idea that a strong trade off leads to specialists, while a weak trade-off promotes generalism, in accordance with the Levins' fitness set approach (Levins 1962, cited in \citealt{RVM06}. 

From the results obtained here it is evident that initial trait values are very important to the final equilibrium of a population. This may be useful in predicting responses when a species is added to or removed from a system, as such an event will be followed by evolution of the system in a direction dictated, to a certain extent, by the equilibrium values of the system prior to the alteration.

One crucial factor concerning this model is that it is dynamical. A structural property can be described and predicted by this dynamical model, explaining an aspect of static models that cannot be explained through a simple trophic hierarchy. This helps in clarifying how parasitic associations may be motivated, and, to a certain extent, investigates the effects which parasite in a host species have on one another. This toy model can now easily be expanded to include a more realistic system containing many more species (in prep.). Analysis of a larger system will then enable questions on the nestedness of the system as a result of trade-offs to be answered.

\subsection{Future}
The primary purpose of this paper was to lay the foundation for a model which could explain patterns of nestedness in ecological networks. In order to do so, this model needs to be repeated at a larger scale for a host-parasite network. Given the evidence for nestedness in other systems \citep{BJMO03,VA03,VA04}, the adaption of this model to these systems, particularly mutualistic networks, could also be used to corroborate any conclusions reached (in prep.). 

Many other factors are considered as possible motivators for the species-richness of parasites in hosts (see \citealt{FRCHDM97,M02,MPS06,MP98,NAJS03,PL11,PM04}), which focus on the characteristics of the hosts. It is becoming increasingly clear that the interests of the parasites are also important factors in this, and, in fact, both motivating factors are likely to be of importance. There have been very few models which investigate the co-evolution of a host-parasite system using adaptive dynamics (but see \citealt{BWB09,CF10}), and these focus primarily on the discovery of a co-evolutionary stable strategy. In that sense, every additional approach to or analysis of a co-evolutionary system adds to the field of co-evolutionary ecology.

With the aim of co-evolution in mind, it has been observed that, according to game theory, predators may be responsible for the presence of additional prey species through induced branching \citep{MB07}. Is it possible that parasites have such an effect? This could presumably only happen if parasites exerted similar levels of pressure to predators, which is unlikely \citep{P10}. Branching in our model could, however, potentially explain nestedness, as parasites would be found in similar hosts. This would lead to results similar to figure \ref{TvsT2a}, an aspect which could be investigated further.

A further step from here is to investigate the effect that the position of a host in the network as a whole has on its parasites, and how this fits in with the observations made here. The position of a host species in the network is considered a potential driving factor in determining its parasitic composition \citep{CLDJHS08,VPKS05}. This has been looked at to a greater extent than parasite interactions with each other in the past \citep{CPD10}, but open questions still remain, specifically with larger networks that include both multiple hosts and multiple parasites together. 

\section{Conclusion}
The results of the model indicate that the hypothesis of resource trade-off driving a link between specificity and species richness appears to be plausible. It can certainly be used to model interactions between hosts and parasites, which should yield interesting results when used on a larger scale. This also highlights the importance of factors such as host mutation rates in co-evolutionary systems, even when these rates are low. 

Using such information as our results for the mutation rates and initial trait values, our model helps to predict the circumstances under which we might expect patterns such as specialization asymmetry to occur. We would predict the presence of specialist parasites in species-rich hosts to be more likely if the hosts had higher mutation rates, and in systems in which parasites are more closely related, are more likely to originate in similar hosts or appear as generalists. Given the relationship between specialization asymmetry and nestedness, we would therefore expect nestedness under similar circumstances, and anti-nestedness the remainder of the time. 

This model demonstrates that dynamic co-evolution of the network is vitally important in accounting for parasites, as it demonstrates how the dynamics could influence structural properties. In particular, it demonstrates the importance of the co-evolution of both hosts and parasites in such a scenario. Parasites, therefore, are not a characteristic to simply be transposed onto a system with no regard to their effects on one-another. Much like interactions in conventional food webs, the influences of different parasites can alter the entire structure of a host-parasite network.
\section*{Acknowledgments}
C.F. McQuaid is a Commonwealth Scholar, funded by the Department for International Development, UK. 

%\nocite{*}
%\bibliographystyle{plain}      % basic style, author-year citations
\bibliographystyle{spmpsci}      % mathematics and physical sciences

\bibliography{Paperbib}

\end{document}